\documentclass[preprint,12pt]{elsarticle}

\usepackage{amssymb}
\usepackage{dcolumn,braket,rotating,afterpage}
\usepackage[version=3]{mhchem}
\usepackage{graphicx}
\usepackage[inline]{enumitem}

\journal{Journal of Luminescence}

\begin{document}

\begin{frontmatter}

	\title{Vacancy Relaxation in Cuprous Oxide (\ce{Cu_{$2-x$}O_{$1-y$}})}

	\author[tu]{Laszlo Frazer\corref{mycorrespondingauthor}\fnref{fn1}}
\cortext[mycorrespondingauthor]{Corresponding author}
\ead{jl@laszlofrazer.com}

\author[chem]{Kelvin B. Chang}
\author[chem,cnm]{Richard D. Schaller}
\author[chem,anl]{Kenneth R. Poeppelmeier}
\author[phys,eecs]{John B. Ketterson}

\fntext[fn1]{Present address:  School of Chemistry, UNSW, Sydney, NSW 2052 Australia.}
\address[tu]{Department of Chemistry and Center for the Computational Design of Functional Layered Materials, 1901 N. 13th Street, Philadelphia, PA 19122, USA}
\address[chem]{Department of Chemistry, Northwestern University, 2145 Sheridan Road, Evanston, IL 60208, USA}
\address[cnm]{Center for Nanoscale Materials, Argonne National Laboratory, 9700 South Cass Avenue, Building 440, Argonne, IL 60439, USA}
\address[anl]{Chemical Sciences and Engineering Division, Argonne National Laboratory, 9700 South Cass Avenue, Argonne, IL 60439, USA}
\address[phys]{Department of Physics, Northwestern University, 2145 Sheridan Road, Evanston, IL 60208, USA}
\address[eecs]{Department of Electrical Engineering and Computer Science, Northwestern University, 2145 Sheridan Road, Evanston, IL 60208, USA}

\begin{abstract}
Phonons are produced when an excited vacancy in cuprous oxide (\ce{Cu2O}) relaxes.  Time resolved luminescence was used to find the excited copper vacancy (acceptor) and oxygen vacancy (donor) trap levels and lifetimes.  It was also used to determine the typical energy and number of phonons in the phonon pulses emitted by vacancies.  The vacancy properties of cuprous oxide are controlled by several synthesis parameters and by the temperature.  We directly demonstrate the absorption of light by oxygen vacancies with transient absorption.  Copper and oxygen vacancies behave differently, in part because the two kinds of traps capture carriers from different states.  For example, the copper vacancy luminescence lifetime is around 25 times greater at low temperature.  However, both kinds of vacancy luminescence are consistent with a Poissonian multiple phonon emission model.
\end{abstract}


\end{frontmatter}

\section{Introduction}
\subsection{Applications of Cuprous Oxide and Importance of its Stoichiometry}
Cuprous oxide (\ce{Cu2O}) is composed of highly abundant elements \cite{isseroff2013electronic,septina2011potentiostatic} and presents a 2.17 eV direct bandgap at room temperature \cite{itoh1975analysis}. Such characteristics make the material a candidate for scalable application in solar energy conversion technologies that range from photovoltaics to photocatalytic water splitting \cite{mcshane2009photocurrent,bendavid2013first,kondo1998cu}. Recent reports also suggest that point defects may aid such technologies by acting as donors or acceptors \cite{mcshane2009photocurrent,wang2015electrodeposited,min2012oxygen,olsen1982experimental}.

Frequently, electronic devices are based on the junction of \emph{p}-type and \emph{n}-type materials.  In cuprous oxide, the type is determined by the nonstoichiometry of the structure and by doping.  The known nonstoichiometries are copper vacancy sites and oxygen vacancy sites.  Cuprous oxide is typically \emph{p}-type because copper vacancies are electron acceptors.  Oygen vacancies, however, are electron donors.  Oxygen vacancies are typically a minority defect.	

In response to experimental reports, Scanlon and Watson have asked, ``Is undoped n-type cuprous oxide fact or fiction?''\cite{scanlon2010undoped}  On a theoretical basis, they conclude that it is fiction.  However, annealing experiments suggest the removal of oxygen vacancies (as opposed to dopants) converts n-type samples to p-type \cite{wang2015electrodeposited,mcshane2009photocurrent}.  There have been several literature reports of geological samples of cuprous oxide which have luminescence from oxygen vacancies, but not from copper vacancies \cite{koirala2013correlated,li2013engineering,ito1997detailed}.  

Cuprous oxide and \ce{Ag2O} are the only materials with achiral octahedral ($O_h$) symmetry.  As a result of the $d$ structure of the valence band and the $s$ structure of the conduction band in these two materials \cite{kazimierczuk2014giant}, exciton radiative decay is suppressed.  Cuprous oxide has a more convenient phase diagram than \ce{Ag2O} \cite{assal1997thermodynamic,schmidt1974growth}.  Therefore, cuprous oxide is a favored material for the investigation of the fundamentals of Wannier-Mott excitons, including the scattering of excitons with other excitons \cite{laszlo2013unexpectedly,wolfe2014search}, phonons \cite{yoshioka2011transition}, and photons \cite{frazer2014photoionization}, or the formation of Bose-Einstein condensates \cite{snoke2014bose,yoshioka2011transition}.  Using samples the same or similar to those reported here, we previously investigated exciton photoionization \cite{frazer2014photoionization} and multiphoton excitation of excitons \cite{frazer2014third}.  Since excitons can decay at vacancies, low vacancy concentration cuprous oxide can be desirable for exciton research.

Here, we use a geological sample and reproducible synthetic samples to investigate whether the exclusive presence of oxygen vacancy luminescence might indicate the sample is \emph{n}-type.  We conclude that the mechanism of relaxation at the two vacancy types is different, so the absence of copper vacancy luminescence does not indicate an absence of copper vacancies.

Trapping and recombination of carriers at defects, which leads to luminescence, reduces the photocurrent in cuprous oxide solar energy devices \cite{isseroff2013electronic}.  Therefore the elimination of vacancy luminescence is desirable.  On the other hand, 
sub-bandgap absorption may be used to capture sunlight which would otherwise be lost in traditional solar energy devices.
The presence of vacancies induces absorption below the bandgap, as we demonstrate below.  

\subsection{Physics of Relaxation in Cuprous Oxide}
Excitation above the bandgap can lead to relaxation into a variety of states, which include (in order of decreasing energy): 
\begin{enumerate*}
	\item electron-hole plasma;
	\item the ``yellow'' exciton series;
	\item excited oxygen vacancy states;
	\item excited copper vacancy states;
	\item excited vibrational states;
	\item the ground state
\end{enumerate*}.
This is not an exhaustive list.  Owing to the high conduction and valence band symmetry, the electron-hole plasma cannot be investigated optically.  Relaxation from the exciton states to vibrational states or the ground state has, however, been studied using luminescence \cite{petroff1975study}.  

As there are a large number of states in the system, the coupling between states can be quite complex.  In this paper, we are measuring relaxation from excited vacancy states to vibrational states.  Since vacancies are fixed at particular lattice site, they are less sensitive to the environment than Wannier-Mott  excitons.  We assume vacancies always relax the same way.  Therefore, variations observed in vacancy relaxation are explained by changes in the behavior of the higher lying states which are causing the vacancy state to become occupied.

\subsection{Structure of this Article}
Section \ref{sec:synth} summarizes the preparation of the samples.
Section \ref{sec:methods} describes how the luminescence was recorded.
Section \ref{sec:spectrum} describes the luminescence spectrum.
Section \ref{sec:dyn} gives the theory of the temporal evolution of the luminescence.
Section \ref{sec:specmodel} gives the theory of the luminescence spectrum.
Section \ref{sec:lifetime} describes control of the luminescence lifetime by different synthesis methods.
Section \ref{sec:energylevel} describes control of the excited energy level of vacancies by different synthesis processes.
Section \ref{sec:phononemission} explains the number and energy of the phonons emitted by vacancies.
Section \ref{sec:power} describes the phonon pulses produced by vacancies.
Section \ref{sec:temp} shows the temperature dependence of the luminescence results.
Section \ref{sec:ta} demonstrates direct absorption by oxygen vacancies using a transient absorption technique.

\section{Material: Cuprous Oxide Synthesis (The Optical Floating Zone Method) \label{sec:synth}}

For this experiment, we selected six samples covering a wide range of synthesis conditions with the goal of demonstrating variability in the results.  As a ``guide to the eye'' throughout this paper, the samples are labeled A to F in order of increasing expected quality.  Expectations were based on our previous measurements of vacancies and inclusions \cite{chang2013removal,frazer2015cupric}. 

Sample A is a natural sample selected for its suspected poor quality.  Most exciton research in cuprous oxide uses naturally mined samples.  A was mined at the Ray Mine, Arizona, USA \cite{clark1998mineral} and purchased from Arkenstone, Richardson, Texas, USA.

Samples B-F were prepared by oxidation, floating zone crystallization, and (excepting C) annealing as previously described \cite{chang2013removal}.  Polycrystalline feed and seed rods of cuprous oxide were obtained through thermal oxidation of copper metal rods.  Oxidation occurred in a furnace at 1045 $^\circ$C for 3 days.  Crystal growth was conducted in an optical image furnace (CSI-FZ-T-10000-H-VI-VP) equipped with four 300 W tungsten halide lamps.  All samples were grown in air.  During growth, the feed and seed rods were counterrotating at 7 RPM.  B, D, E, and F were annealed post growth at 1045 $^\circ$C. The samples were polished on two parallel faces with a process discussed elsewhere \cite{frazer2014excitons}.  The detailed synthesis parameters are listed in Table \ref{tab:samples}.

To summarize our previous results, we initially compared the intensity of copper vacancy luminescence across samples \cite{chang2013removal}.  Images of B, C, and D are shown in Figure 8 of \cite{chang2013removal}. We assumed that the luminescence intensity increases with the copper vacancy concentration.  The vacancy luminescence is the most sensitive qualitative probe of stoichiometry available.  
\begin{sidewaystable}
\begin{tabular}{rlllllll}
	Symbol  &Thickness (mm) &Purity &AD (days) &CR (C/minute)   &Power (\%)     &GR (mm/hour)    &Diameter (mm)\\\hline
	A       &2.0    &Geological     &       &       &       &\\
	B       &0.61   &0.999  &4      &Quench &55.4   &7      &5\\
	C       &2.3    &0.999  &       &       &55.4   &7      &5\\
	D       &1.8    &0.999  &4      &5      &55.4   &7      &5\\
	E       &0.74   &0.9999 &7      &5      &56.5   &3.5    &6.35\\
	F       &0.24   &0.99999&3      &5      &55.8   &3.5    &5\\
\end{tabular}
\caption{Cuprous oxide synthesis parameters.  Purity:  Refers to copper precursor purity. AD:  Anneal duration. CR:  Cooling rate; ramp rate of the furnace during annealing.  Power:  Floating zone furnace lamp power.  GR:  Growth rate.  \label{tab:samples}}
\end{sidewaystable}

\section{Experimental: Luminescence Methods\label{sec:methods}}
We used a 35 fs regeneratively amplified titanium doped sapphire laser.  The laser output was converted to 500 nm using an optical parametric amplifier and a $\beta$-barium borate second harmonic generator.  We selected 500 nm as the excitation wavelength because it is comparable with wavelengths widely used in the literature.  Previous experiments have used continuous wave argon-ion and intracavity doubled neodymium-based excitation lasers.  We used 0.4 to 1 mW of laser power to excite each sample.  The laser pulsed at a repetition rate of 2 kHz.  The beam diameter was about 1 mm.  The peak irradiance was about 10$^{13}$ Watts/m$^2$.

The samples were mounted with VGE-7031 varnish on a sapphire window.  The sample assembly was placed in vacuum in a cryostat held at 3.2 Kelvin, except where other temperatures are specified below.  The luminescence was collected in a transmission geometry.  We used two detectors.  
The first was a fiber coupled, 300 mm focal length spectrograph.  The spectrograph was equipped with 150 and 1800 groove/mm gratings. These were used for vacancy and exciton luminescence, respectively.  The spectrograph data was recorded with a CCD.  
The second detection system was for time resolved luminescence.  It included a 150 mm focal length spectrograph with a 50 groove/mm grating.  The time resolved spectra were recorded using a Hamamatsu streak camera operating in single photon counting mode.  Luminescence intensities were not intended to be comparable across samples.  The efficiency of spectrometers is not constant over wide ranges of the spectrum.

\begin{figure}
	\begin{center}
	\includegraphics[width=.7\textwidth]{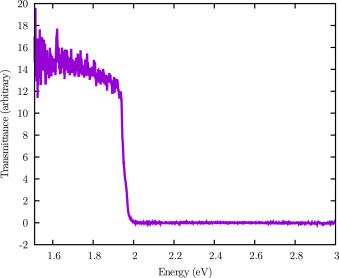}
	\end{center}
	\caption{Room temperature transmittance spectrum of sample F.\label{fig:trans}}
\end{figure}

The absorption spectrum of cuprous oxide is not known to have any features in the region of the spectrum where the vacancy luminescence is present.  Fig. \ref{fig:trans} shows the transmittance spectrum of sample F at room temperature over a limited range.  The transmittance below the bandgap is primarily impacted by the high reflectivity of the crystal.  At room temperature, the absorption band edge is broadened by indirect transitions \cite{baumeister1961optical}.  The absorption of photoluminescence by the sample does not significantly modify the photoluminescence spectrum.

\section{Experimental: Time Averaged Luminescence Spectroscopy\label{sec:spectrum}}
\begin{figure}
	\begin{center}
	\includegraphics[width=.7\textwidth]{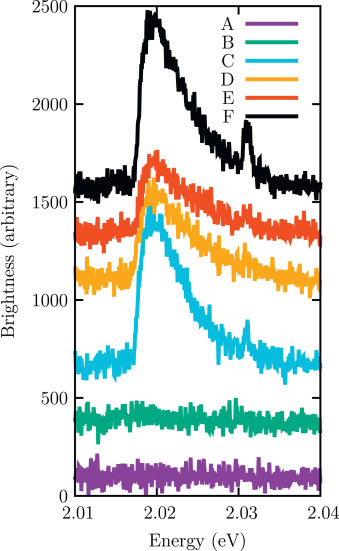}
	\end{center}
	\caption{Exciton luminescence spectra from the indicated samples.  The temperature is 3.2 K. The broad peak at 2.02 eV is caused by decay of a \emph{1s} orthoexciton into a 13.8 meV $^2\Gamma_{12}^-$ phonon and photon.  The sharp peak at 2.032 eV is the quadrupole \emph{1s} orthoexciton polariton.  A and B lack significant exciton luminescence.  In high quality samples, fewer excitons are trapped in defects, so more excitons decay in the form of polaritons.  In addition, in high quality samples the excitons thermalize more before decaying, so the phonon assisted peak becomes narrower.  The spectra are offset for clarity.\label{fig:exciton}}
\end{figure}

There are five luminescence peaks we consider important in our luminescence spectra.  These peaks are previously reported.  They will be described systematically from highest to lowest energy.
Cuprous oxide has a large number of exciton states \cite{kazimierczuk2014giant,schmutzler2013signatures}.  The conduction and valence band edges have positive parity symmetry.  Since light has negative parity, coupling of the band edge transitions or exciton to light is suppressed in cuprous oxide.  The primary source of direct free exciton luminescence is quadrupole polariton formation \cite{jang2008indirect}.  This occurs only for the \emph{1s} orthoexciton polariton state.  This state is the second excited exciton state, since it is slightly above the paraexciton state \cite{frohlich2005high,kubouchi2005study}.  Paraexcitons are not detected here.  In Figure \ref{fig:exciton}, the exciton luminescence at 2.03 eV corresponds to orthoexciton polariton luminescence.  

The orthoexciton can also decay into one of several phonons and a photon \cite{petroff1975study}.  This is known as a phonon-assisted or Stokes process.  It appears most strongly at 2.02 eV in Figure \ref{fig:exciton}.  This peak corresponds to emission of a $^2\Gamma_{12}^-$ phonon.  The phonon corresponds to compression of opposing edges of the copper tetrahedra surrounding each oxygen atom \cite{elliott1961symmetry,oharathesis}.

\label{}

The phonon-assisted luminescence spectrum adopts a Maxwell-Boltzmann distribution \cite{frazer2015evaluation,gross1966free}.  This is because the spectrum is a convolution of the exciton distribution and the phonon density of states.  However, the optical phonon density of states is approximately monochromatic.  In other words, the phonon effective mass is large \cite{korzhavyi2011literature}.  Therefore the phonon-assisted luminescence allows us to read out the exciton energy distribution.

If $E_c$ is the threshold energy at which the luminescence starts, $T$ is the exciton temperature, and $k_B$ is the Boltzmann constant, then the luminescence as a function of energy is
\begin{align}
	I(E)&=A\left(\left|E-E_c\right|\right)^{\frac12}e^{-\frac{E-E_c}{k_BT}}.\label{mb} 
\end{align}
We used this equation to determine the temperature of the excitons \cite{frazer2015evaluation,gross1966free} where sufficient brightness was collected.  Exciton temperatures are generally higher than crystal lattice temperatures because excitons are not in thermal equilibrium with their environment \cite{snoke1989population}.

\begin{figure}
	\begin{center}
	\includegraphics[width=.7\textwidth]{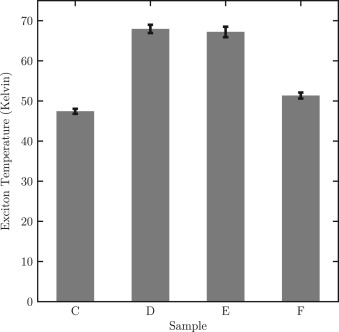}
	\end{center}
	\caption{Exciton temperature determined from a Maxwell-Boltzmann model of the phonon-assisted luminescence.  Samples A and B did not produce exciton luminescence.\label{fig:excitontemp}}
\end{figure}

We found that samples A and B did not show any exciton luminescence.  Both samples contain visible impurities.  Sample F has the strongest exciton luminescence.  The phonon-assisted peak is noticeably sharper than the peaks for samples D and E, which are lower purity.  Longer exciton lifetimes lead to greater exciton thermalization with the lattice.  This leads to a lower exciton temperature, and a narrower phonon-assisted peak.  The exciton temperatures shown in Figure \ref{fig:excitontemp} are negatively correlated with the copper vacancy luminescence lifetimes, as will be shown in Table \ref{tab:results}.

\begin{figure}
	\begin{center}
	\includegraphics[width=.7\textwidth]{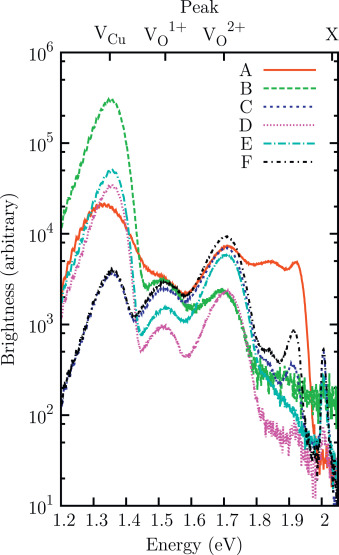}
	\end{center}
	\caption{Time averaged vacancy luminescence spectra for the samples.  The luminescence peaks are labeled at the top.  From left to right, they are copper vacancy luminescence (\ce{V$_{\text{Cu}}$}), singly ionized oxygen vacancy luminescence (\ce{V$_{\text{O}}^{1+}$}), doubly ionized oxygen vacancy luminescence (\ce{V$_{\text{O}}^{2+}$}), and phonon-assisted exciton luminescence (X).  We do not identify the extrinsic feature at 1.9 eV.\label{fig:vaclumin}}
\end{figure}

Figure \ref{fig:vaclumin} shows wider field of view, lower resolution luminescence spectra of the samples.  In this case, the exciton and phonon-assisted exciton peaks are not distinguished from each other.  Between 1.8 and 1.95 eV, some samples, and especially A, show luminescence which we have not identified.  The vacancy luminescence peaks are broad.  The luminescence from a doubly ionized oxygen vacancy (\ce{V$_{\text{O}}^{2+}$}) is greatest near 1.7 eV.  A smaller peak, which scales proportionately, is  caused by singly ionized oxygen vacancies (\ce{V$_{\text{O}}^{1+}$}).  In previous experiments, the singly ionized oxygen vacancy peak has been reported inconsistently \cite{markworth2001epitaxial,ito1998single,ito1997detailed}.  We suggest this may be caused by variations in detection geometry.   Our previous reflection-geometry spectra of some of the same samples did not show \ce{V$_{\text{O}}^{1+}$} peaks clearly \cite{frazer2015evaluation}.

The copper vacancy (\ce{V$_{\text{Cu}}$}) luminescence peak is brightest near 1.35 eV.  In previous reports, copper vacancy luminescence has predominated in synthetic samples.  Only oxygen vacancy luminescence was present in geological samples in some literature reports \cite{koirala2013correlated,li2013engineering,ito1997detailed}.  All our samples show both types of luminescence, in varying relative amounts.  The relative intensities of the time averaged luminescence from the different peaks is known to be temperature dependent \cite{ito1997detailed,li2013engineering}.

\section{Time Resolved Luminescence Spectroscopy}
\begin{figure}
	\begin{center}
	\includegraphics[width=\textwidth]{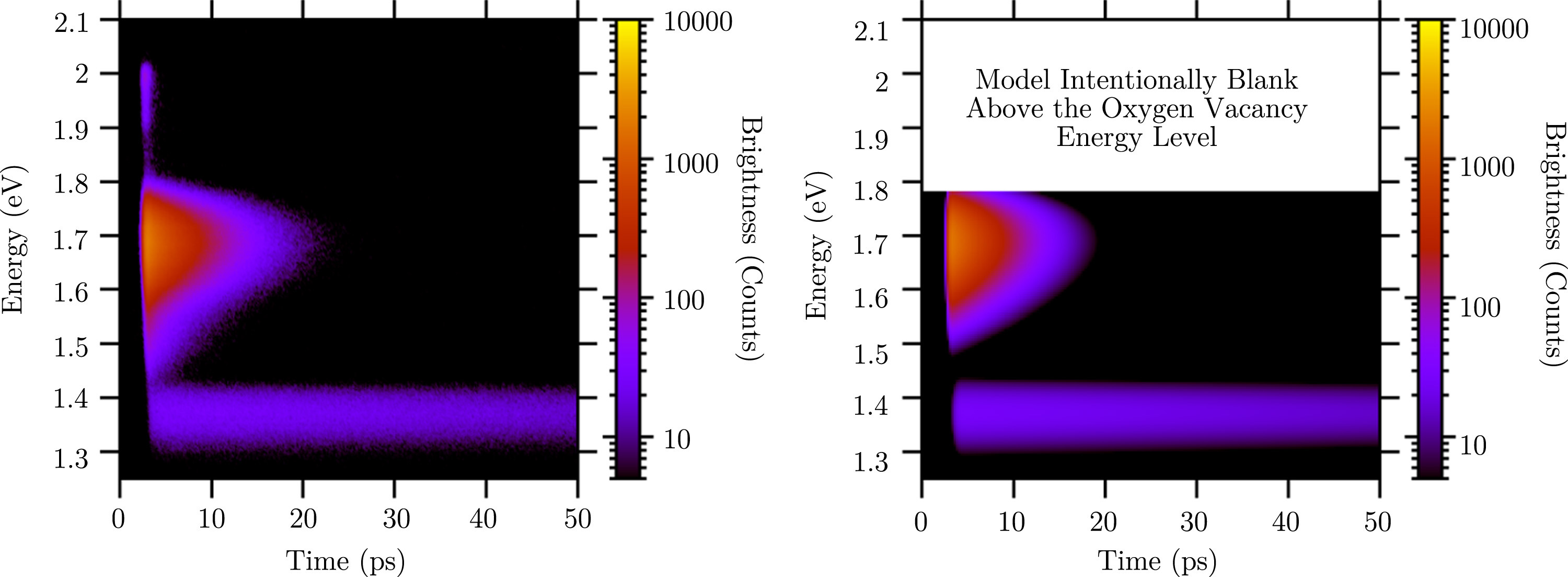}
	\end{center}
	\caption{Left:  Experimental time resolved vacancy luminescence for sample D.  Right:  Model of the luminescence.  The model does not apply above 1.78 eV, so that area is left blank. The color scale is logarithmic to clearly show both the oxygen and copper vacancy luminescence. \label{fig:example}}
\end{figure}

We model both the spectrum and the dynamics of the time resolved vacancy luminescence.  Luminescence sources at energies above the oxygen vacancy energy level, such as exciton luminescence, were not modeled and have been studied elsewhere \cite{frazer2015evaluation,laszlo2013unexpectedly,shen1997dynamics}.  Figure \ref{fig:example} shows an example of time resolved luminescence and our model. These are in good agreement.  

\subsection{Theory: Dynamical Model\label{sec:dyn}}
We model the time dependence of the vacancy luminescence with two physical processes.  The first one is an initiating pulse.  This pulse could be interpreted as the laser pulse exciting the vacancy directly, or it could be interpreted as an approximation of a more complex process.  Vacancy luminescence may be preceded by several steps, including the formation of electron-hole plasma and the formation of free excitons.

The second process is decay of the luminescence.  This can be caused by the intrinsic lifetime of an exciton bound to a vacancy.  Alternatively, it could reflect the lifetime of the state pumping the vacancy state.  Reference \cite{koirala2013correlated} finds the oxygen vacancy lifetime is equal to the exciton lifetime below 40 Kelvin.  In that experiment, little copper vacancy luminescence was observed.  Our results will show that in samples with both vacancy types, the physics is different.

A pump pulse with amplitude $A$ and duration $\sigma$ excites the vacancies.  The lifetime of the excited vacancy population is $\tau$.  In a prompt excitation model, the number of excited vacancies $N$ at time $t$ evolves according to the model
\begin{align}
	\frac{dN}{dt}&=A e^{-\frac{t^2}{2\sigma^2}}-\frac{N}{\tau}
	\label{decayrate}.
\end{align}
The solution to the differential equation is
\begin{align}
	N(t)&=Ae^{-\frac{t}{\tau}}e^{\frac{\sigma^2}{2\tau^2}}\sigma \sqrt{\frac{\pi}{2}} \left( 1-\operatorname{erf}\left(\frac{\sigma^2-\tau t}{\sqrt2\tau\sigma}  \right) \right)
	\label{dynamics}
\end{align}
where $\operatorname{erf}$ is the error function.  

\subsection{Theory: Multiple Phonon Emission Spectral Model\label{sec:specmodel}}
Based on Reference \cite{koirala2014relaxation}, excitons decay at oxygen vacancies by the emission of multiple phonons.  Each phonon emitted subtracts some energy from the luminescence produced by the vacancy.  Zero-phonon lines are widely reported in other materials, but we are not aware of any evidence for zero-phonon lines from vacancies in cuprous oxide.  

Using the Poisson distribution and the assumption that the phonon modes initially have zero occupation \cite{koirala2014relaxation}, for $n$ phonons emitted and a Huang-Rhys factor $S$, the amount of luminescence is
\begin{align}
	I(n,S)&\propto\frac{e^{-S}S^n}{n!}.\label{oldmodel}
\end{align}
This model agrees with the observed spectrum \cite{koirala2014relaxation}.  However, individual peaks for specific $n$ are not observed for several reasons:
\begin{enumerate*}
	\item finite lifetime \cite{koirala2014relaxation};
	\item inhomogeneous broadening \cite{koirala2014relaxation};
	\item vacancies may modify the phonon spectrum \cite{koirala2014relaxation};
	\item there are eight distinct phonon energies at zone center \cite{petroff1975study}.  A mixture of these modes may be emitted.  In the red luminescence of brown diamonds, there is a dominant, high-energy mode for which $n$ phonon emission peaks can be resolved at low temperature \cite{pereira1986red};
	\item our streak camera does not have sufficient spectral resolution to distinguish closely spaced peaks.
\end{enumerate*}
For these reasons, we have created a smoothed spectral model by replacing the discrete factorial function with the continuous $\Gamma$ function.  The luminescence at energy $E$, for vacancy energy level $E_V$ and typical phonon energy $E_p$ is
\begin{align}
	I(E,S)&\propto\frac{e^{-S}S^{\frac{E-E_V}{E_p}}}{\Gamma\left({\frac{E-E_V}{E_p}}+1\right)}.
	\label{spectrum}
\end{align}
$E_p$ is not associated with a particular phonon mode, and may reflect a mixture of phonon types emitted.  The Huang-Rhys factor $S$ represents the typical number of phonons emitted during each vacancy luminescence event.

Model \ref{spectrum} interpolates Model \ref{oldmodel}.  For nonnegative integer values of $(E-E_V)/E_p$ the two models are identical \cite{gamma}.  When least-squares fit to the spectra, the parameters of the model converge to unique values because the relationships between the variables and the parameters are independent.

\subsection{Results: Luminescence Lifetime\label{sec:lifetime}}
Reference \cite{koirala2014relaxation} did not report copper vacancy luminescence.  We have found that the modified Poissonian model \ref{spectrum} is consistent with the copper vacancy luminescence as well as the \ce{V$_{\text{O}}^{2+}$} luminescence.  The two vacancies are readily distinguished by their different energy levels $E_V$.  We did not model the weaker \ce{V$_{\text{O}}^{1+}$} luminescence, which may skew the oxygen vacancy model slightly.  

The dynamical and spectral parameters from Equations \ref{dynamics} and \ref{spectrum} were simultaneously fit with separate parameters for each of the two luminescence peaks.  For half the samples, there was not enough copper vacancy luminescence for this global analysis approach to work.  For C and F, copper vacancy luminescence lifetimes were determined using spectrally integrated brightness.  For A, no copper vacancy luminescence was detected.  For copper vacancy luminescence from sample C, only about one photon per 100 fs time bin was counted in the data available.  Therefore, it was necessary to include a background term in the model to account for the dark counts in this case.

\begin{figure}
	\begin{center}
	\includegraphics[width=.7\textwidth]{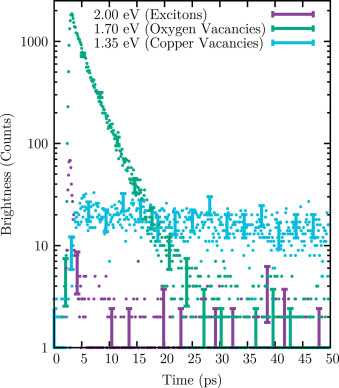}
	\end{center}
	\caption{Time resolved luminescence data from Figure \ref{fig:example} at particular energies.  The sample is D.  For clarity, only selected error bars are shown.  The exciton luminescence is phonon-assisted.\label{fig:examplecut}}
\end{figure}

Figure \ref{fig:examplecut} shows the lifetime of the copper vacancy luminescence is much greater than the lifetime of the oxygen vacancy luminescence.  In our view, the long lifetime of the copper vacancy luminescence is caused by long lived free excitons.  The luminescence continues for the lifetime of the excitons because free excitons can decay by binding to the vacancies.   An alternative hypothesis is that this is the intrinsic lifetime of the vacancy luminescence.

Table \ref{tab:results} shows that for samples D, E, and F, both types of luminescence lifetime increase.  These samples were synthesized from copper precursors with increasing purity.  Certain impurities in the host metal may catalyze vacancy formation \cite{isseroff2013electronic,chang2013removal}.  Fewer vacancies lead to longer free exciton lifetimes.  Since the excitons pump the vacancies, this in turn leads to a longer luminescence lifetime.
\begin{sidewaystable}
\begin{tabular*}{.48\textwidth}{rD{,}{\ \pm\ }{5}D{,}{\ \pm\ }{9}D{,}{\ \pm\ }{5}D{,}{\ \pm\ }{4}D{,}{\ \pm\ }{6}D{,}{\ \pm\ }{5}}
	Sample&\multicolumn{1}{c}{\ce{V$_{\text{O}}^{2+}$}: $\tau$ (ps)}&\multicolumn{1}{c}{\ce{V$_{\text{Cu}}$}: $\tau$ (ps)}&\multicolumn{1}{c}{\ce{V$_{\text{O}}^{2+}$}: $E_V$ (eV)}&\multicolumn{1}{c}{\ce{V$_{\text{O}}^{2+}$}: $S$}&\multicolumn{1}{c}{\ce{V$_{\text{O}}^{2+}$}: $E_p$ (eV)}&\multicolumn{1}{c}{\ce{V$_{\text{O}}^{2+}$}: $P$ (nW)}\\\cline{1-7}
A&4.013,0.003&&1.7847,0.0004&4.55,0.03&0.02414,0.00008&4.39,0.03\\
B&2.967,0.004&83.3,0.3&1.7611,0.0005&3.46,0.03&0.0316\phantom{0},0.0002&5.91,0.06\\ 
C&3.107,0.002&(1.6,0.2)\times 10^2&1.7869,0.0003&5.00,0.03&0.02230,0.00006&5.75,0.03\\
D&2.705,0.001&81.4,0.8&1.7839,0.0002&4.38,0.02&0.02410,0.00005&6.25,0.03\\
E&3.010,0.002&91,1&1.7866,0.0004&4.44,0.03&0.02396,0.00009&5.66,0.04\\
F&3.394,0.002&(1.0,0.3)\times 10^2&1.7858,0.0003&4.29,0.02&0.02406,0.00007&4.87,0.03\\
Ref. \cite{koirala2014relaxation}&&&1.833\phantom{0}&7.9\phantom{0},0.3&0.0161\phantom{0},0.0003&
\end{tabular*}
	\caption{Results from modeling the time resolved luminescence of the samples.
\ce{V$_{\text{O}}^{2+}$}: $\tau$;
	Lifetime of oxygen vacancy luminescence.
\ce{V$_{\text{Cu}}$}: $\tau$;
	Lifetime of copper vacancy luminescence.
\ce{V$_{\text{O}}^{2+}$}: $E_V$;
	Energy level of the oxygen vacancies obtained from the time resolved luminescence.
\ce{V$_{\text{O}}^{2+}$}: $S$;
	Huang-Rhys Factor of the oxygen vacancies obtained from the time resolved luminescence. 
\ce{V$_{\text{O}}^{2+}$}: $E_p$;
	Typical energy of the phonons emitted during relaxation at oxygen vacancies obtained from the time resolved luminescence.
\ce{V$_{\text{O}}^{2+}$}: $P$;
	The oxygen vacancy phonon power obtained from time resolved luminescence. 
	\label{tab:results}}
\end{sidewaystable}

The excitation of oxygen vacancies is prompt.  The excitation of copper vacancies is delayed because it is triggered by the trapping of mobile excitons.  For the oxygen vacancy luminescence, the rise time $\sigma$ was always consistent with the instrument resolution, about 0.2 ps.  For the copper vacancy luminescence, the rise time ranged up to 0.9 ps.  We expect that this rise time is temperature and laser power dependent.  It is based on the exciton diffusion constant, which has previously been investigated \cite{trauernicht1984highly}.  

\subsection{Results: Vacancy Energy Level\label{sec:energylevel}}

The energy level of the excited oxygen vacancy varies slightly, but statically significantly, between samples in Table \ref{tab:results}.  
The calculated uncertainties use estimates of the noise in the measurement, indicating the difference between samples is a sample property and not based on brightness noise.
Sample B is  substantially different from the others.  Since B has a high concentration of inclusions, which cause a large strain field \cite{frazer2015cupric}, the results suggest that strain may lower the energy level.  Our higher quality samples, D, E, and F, show highly consistent results.  The copper vacancy energy level is 1.5 eV for samples B, D, and E.  The energy levels calculated by density functional theory are reasonably accurate for copper vacancies but are underestimates for oxygen vacancies \cite{scanlon2010undoped}.

In Table \ref{tab:results}, we also show the oxygen vacancy properties reported in Reference \cite{koirala2014relaxation}.  Like sample A, this data is from a geological sample.  The results and methods have some minor differences from this report.

\subsection{Results: The Nature of Phonon Emission\label{sec:phononemission}}
Based on the measured Huang-Rhys factors shown in Table \ref{tab:results}, in cuprous oxide each oxygen vacancy decay event involves the emission of about four phonons.  In zinc oxide, the Huang-Rhys factor for the green band is 6.5 \cite{shi2006green}.  This zinc oxide band originates from oxygen vacancies \cite{vanheusden1996mechanisms}.  Table \ref{tab:results} shows the typical energy of the phonons emitted is 24 meV. The higher quality samples, D, E, and F, show consistent results.  For copper vacancies in samples B, D, and E, the Huang-Rhys factor is eight to eleven phonons per vacancy per event.  The typical phonon energy is just 14 meV.

\subsection{Results: Vacancy Phonon Power\label{sec:power}}
The typical power output, in the form of phonons, from each single vacancy relaxation is 
\begin{align}
	P&=\frac{SE_p}{\tau}.
\end{align}
The oxygen vacancy phonon power is shown in Table \ref{tab:results}.  Based on samples B, D, and E, the copper vacancy phonon power is 0.2 to 0.3 nW.  

A few phonons are emitted very quickly --- on the picosecond scale.  Copper vacancies emit the phonons somewhat more slowly because they take longer to become excited.  Previously, we considered vacancies detrimental to the formation of Bose-Einstein condensates of excitons in cuprous oxide because the vacancies removed excitons from the system.  These results show that phonon-emitting vacancies can, in addition, heat the excitons by phonon absorption \cite{yoshioka2011transition}, which can keep the excitons above the transition temperature.
\section{Results: Temperature Dependent Time Resolved Luminescence\label{sec:temp}}

We further investigated the temperature dependence of the luminescence from Sample D. D was selected because it was prepared with our recommended inclusion removing technique \cite{chang2013removal,frazer2015cupric}, but shows relatively more vacancy luminescence because of precursor impurities.  Owing to the weakness of the copper vacancy luminescence at higher temperatures, only the dynamics of the copper vacancies was analyzed.  For the oxygen vacancies, we were able to perform global analysis of the dynamics and spectrum. 

\begin{figure}
	\begin{center}
	\includegraphics[width=.6\textwidth]{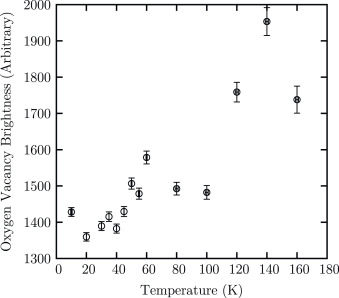}
	\end{center}
	\caption{Oxygen vacancy brightness as a function of temperature.  In contrast to the time-averaged spectra, the time resolved oxygen vacancy luminescence gets brighter with increasing temperature.  The sample is D.\label{fig:oamptemp}}
\end{figure}

\begin{figure}
	\begin{center}
	\includegraphics[width=.6\textwidth]{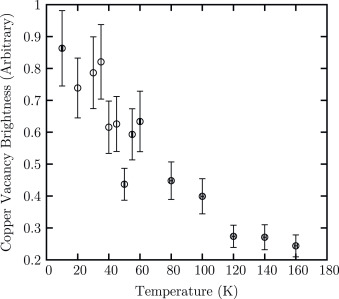}
	\end{center}
	\caption{Copper vacancy brightness as a function of temperature.  Copper vacancy luminescence decreases with increasing temperature.  The sample is D.\label{fig:camptemp}}
\end{figure}

Time averaged, temperature dependent luminescence previously showed that the behavior of vacancies is strongly temperature dependent \cite{ito1997detailed}.  Both kinds of luminescence decrease with temperature.  The oxygen vacancy luminescence is not detected at higher temperatures.  In contrast with the time averaged data, the temperature dependent luminescence in Figure \ref{fig:oamptemp} shows that the amplitude parameter $A$ increases with temperature in the case of oxygen vacancies.  The copper vacancy luminescence $A$ decreases with temperature in Figure \ref{fig:camptemp}.  

\begin{figure}
	\begin{center}
	\includegraphics[width=.6\textwidth]{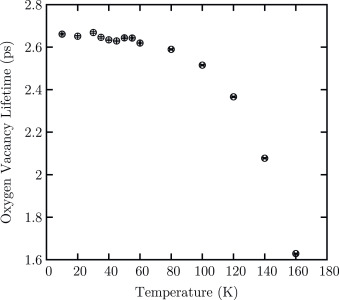}
	\end{center}
	\caption{Oxygen vacancy lifetime as a function of temperature.  Oxygen vacancy lifetime decreases with temperature.  The sample is D.\label{fig:olifetimetemp}}
\end{figure}

\begin{figure}
	\begin{center}
	\includegraphics[width=.6\textwidth]{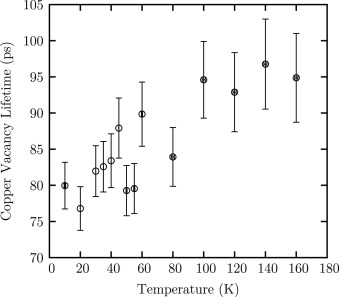}
	\end{center}
	\caption{Copper vacancy lifetime as a function of temperature.  Usually, relaxation is faster at higher temperatures.  That is not the case here.  The copper vacancy relaxation slows down because phonon scattering decreases the exciton diffusion constant.  This increases the time required for excitons to be captured at a copper vacancy.  The sample is D.\label{fig:clifetimetemp}}
\end{figure}

If the luminescence lifetime is considered in addition to the brightness, the time resolved data are in agreement with the time averaged results.   Figure \ref{fig:olifetimetemp} shows that the oxygen vacancy luminescence lifetime $\tau$ plummets at higher temperatures.  However, in Figure \ref{fig:clifetimetemp} the copper vacancy luminescence lifetime increases slightly.  Therefore, the decrease in time averaged vacancy luminescence observed with increasing temperature is dominated by the dynamics of the vacancy luminescence.  

\begin{figure}
	\begin{center}
	\includegraphics[width=.6\textwidth]{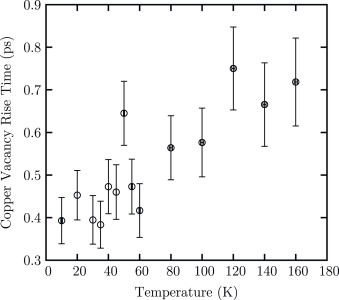}
	\end{center}
	\caption{The rise time of copper vacancy luminescence as a function of temperature.  Exciton capture is impeded by exciton-phonon scattering.   The sample is D.\label{fig:crisetemp}}
\end{figure}

Higher lattice temperatures lead to higher steady state phonon concentrations.  The thermal phonons have two roles.  They accelerate thermalization \cite{snoke1989population,sobkowiak2014modeling,snoke1991carrier}.  They also slow down diffusion of thermalized excitons \cite{trauernicht1984highly}.  The temperature dependent lifetimes show that the copper vacancies are excited primarily by thermalized, diffusive excitons.  This non-prompt interpretation is further supported by the increase in the copper vacancy rise time $\sigma$ with temperature in Figure \ref{fig:crisetemp}.  Thermal phonons are preventing excitons from reaching the copper vacancies before time $\sigma$.
The distinctive behavior of the oxygen vacancies may be explained if they are preferentially excited by nonthermalized excitons or electrons from the electron-hole plasma.

\begin{figure}
	\begin{center}
	\includegraphics[width=.6\textwidth]{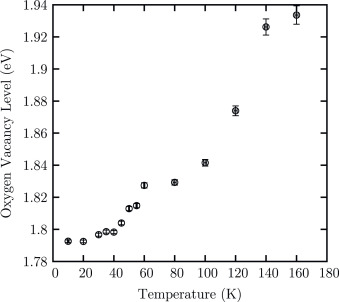}
	\end{center}
	\caption{Oxygen vacancy energy level as a function of temperature, showing the level rises with temperature.  The sample is D.\label{fig:oleveltemp}}
\end{figure}

\begin{figure}
	\begin{center}
	\includegraphics[width=.6\textwidth]{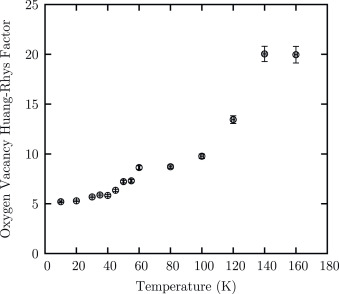}
	\end{center}
	\caption{Oxygen vacancy Huang-Rhys Factor as a function of temperature, showing the Huang-Rhys Factor increases with temperature.  The sample is D.\label{fig:ohrtemp}}
\end{figure}

\begin{figure}
	\begin{center}
	\includegraphics[width=.6\textwidth]{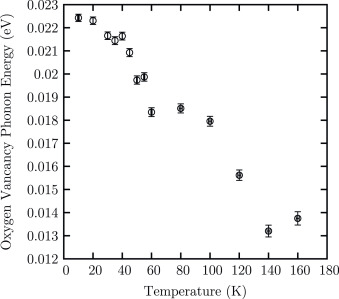}
	\end{center}
	\caption{$E_p$ as a function of temperature. The typical energy of the phonons emitted during relaxation at oxygen vacancies decreases with temperature.  The sample is D.\label{fig:ophonontemp}}
\end{figure}

\begin{figure}
	\begin{center}
	\includegraphics[width=.6\textwidth]{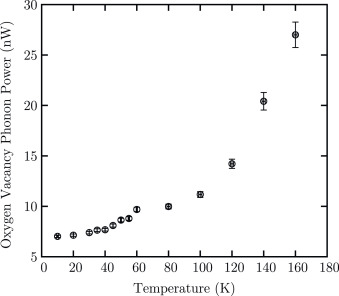}
	\end{center}
	\caption{The oxygen vacancy $P$ as a function of temperature.  The oxygen vacancy phonon power increases with temperature.  The sample is D.\label{fig:opowertemp}}
\end{figure}

In Figure \ref{fig:oleveltemp}, the energy level of the oxygen vacancy increases with temperature, similar to the shift in the energy of the bandgap \cite{itoh1975analysis}.  The number of phonons emitted during each luminescence event (Figure \ref{fig:ohrtemp}) increases with temperature.  However, the typical energy of each phonon emitted declines (Figure \ref{fig:ophonontemp}).  Figure \ref{fig:opowertemp} shows that the net effect is a large increase in the phonon power of oxygen vacancies.  On the whole, the results suggest that while oxygen vacancies are a poor source of time averaged luminescence at room temperature, they are are a good source of photoexcited, low energy phonons.

One might expect the relaxation of a defect state which is far above the valence band relative to the thermal energy $\frac12k_BT$ to be temperature-insensitive.  Our observations that the vacancy dynamics, Huang-Rhys factor, and typical phonon energy change with temperature for the oxygen vacancy explain the steady state spectra.  However, the underlying theory of excited vacancy/phonon coupling remains undetermined.  We speculate that the structure of the oxygen vacancy may be temperature dependent.



\clearpage 

\section{Transient Absorption at Oxygen Vacancies\label{sec:ta}}
Based on previous laser absorption-luminescence spectroscopy \cite{koirala2014relaxation}, we expect that oxygen vacancies absorb light at energies between the vacancy energy level and the intrinsic semiconductor absorption edge.  The mechanism for absorption is transfer of an electron from the valence band to the vacancy plus emission of multiple phonons.  Owing to the low concentration of oxygen vacancies, the vacancy absorption was not directly observed in previous studies \cite{koirala2014relaxation}.  Therefore, we performed a transient absorption measurement where the absorption change caused by excitation of vacancies was measured.  
\subsection{Experimental: Transient Absorption}
The sample was excited at 400 nm using the laser second harmonic.  The excitation beam was focused on to the sample in a spot with a diameter of 0.85 mm to $1/e$.  The absorption was probed using white light produced by self phase matching.  The probe was recorded using an Ultrafast Systems Helios spectrometer.  The temperature was 3.2 Kelvin.  This instrument cannot measure in the energy region where copper vacancies are expected to absorb.
\subsection{Results}
\begin{figure}
	\begin{center}
	\includegraphics[width=.7\textwidth]{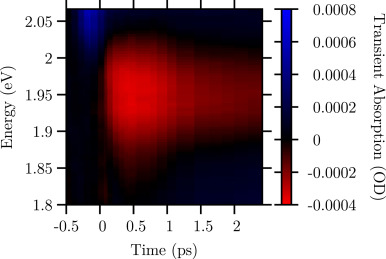}
	\end{center}
	\caption{Transient absorption ``bleach'' of the oxygen vacancy absorption. The sample is crystal F.  The pump radiant exposure is 1.6 J/m$^2$. \label{fig:ta}}
\end{figure}

The absorption by the oxygen vacancies is reduced when they are excited.  The optical density (OD) of this ``bleaching'' is shown in Figure \ref{fig:ta}.  It is strongest between 1.9 and 2.0 eV.  At higher energies, intrinsic absorption \cite{baumeister1961optical} is expected to prevent the observation of bleaching.  The multiple phonon emission model of the induced transparency spectrum does not agree with the data, as expected \cite{koirala2014relaxation}.  The reason may be that in absorption, the number of phonons produced is determined by the fixed energy of the absorbed light and the vacancy state. In contrast, in luminescence there is a Poissonian statistical process because the energy of the spontaneously produced photon is not fixed.  We also observed the weak coherent artifact typically found in transient absorption experiments \cite{dietzek2007appearance,kovalenko1999femtosecond}.  

\begin{figure}
	\begin{center}
	\includegraphics[width=.7\textwidth]{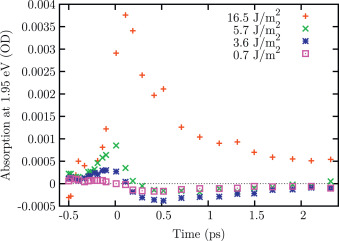}
	\end{center}
	\caption{The negative transient absorption at oxygen vacancies is best detected at lower excitation radiant exposures (about 4 J/m$^2$).  Otherwise it is overwhelmed by induced absorption.  The sample is crystal F.  \label{fig:bleachpower}}
\end{figure}

There is also a broadband induced absorption.  Suppressing the induced absorption is critical to reproducing the data.  If the induced absorption is caused by electron-hole plasma, it will increase as the pump laser radiant exposure increases.  However, the oxygen vacancy bleach cannot increase any more once all the vacancies are excited; we expect saturation of the vacancy states.  In Figure \ref{fig:bleachpower}, we find that the bleaching increases up to about 4 J/m$^2$.  At higher radiant exposures, it is overwhelmed by the induced absorption.

\begin{figure}
	\begin{center}
	\includegraphics[width=.7\textwidth]{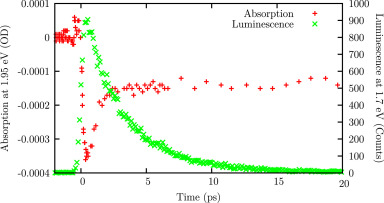}
	\end{center}
	\caption{Overlay of the transient absorption at 1.95 eV from Figure \ref{fig:ta} and the time resolved luminescence at 1.7 eV.  The absorption recovers faster than the luminescence, but does not return to baseline. The sample is crystal F.  \label{fig:dynamics}}
\end{figure}

The bleaching of the oxygen vacancy absorption cannot last longer than the luminescence.  However, it could be shorter because there is an intermediate state which occurs between absorption and luminescence.  Direct comparison of the luminescence and the absorption shown in Figure \ref{fig:dynamics} suggests this is the case.  

The approximately 3 ps luminescence lifetime observed in Figure \ref{fig:olifetimetemp} does not vary much at low temperature.  Based on the discrepancy between the bleach lifetime and the luminescence lifetime, we expect the low temperature luminescence lifetime in the oxygen vacancy is the inherent lifetime of the vacancy excited state, as opposed to the lifetime of a higher lying state which is driving excitation of the vacancy.  This non-prompt interpretation is further support for the concept that oxygen vacancies are preferentially excited by a higher lying state than the copper vacancies.

Finally, we note that the bleach does not fully recover in Figure \ref{fig:dynamics}.  The long lived portion of the bleach would be an interesting area of future investigation.

\afterpage{\clearpage}
\section{Conclusions}
The relaxation luminescence, whether it comes from copper vacancies or oxygen vacancies, is consistent with multiple phonon emission.  However, the physical parameters for the two vacancy types are different in every case.
More, lower energy phonons are emitted during copper vacancy relaxation than during oxygen vacancy relaxation.
The number of phonons emitted during relaxation at vacancies increases with temperature and can be modified by defects.  
The typical energy of phonons emitted during relaxation at vacancies decreases with temperature and can also be modified by defects.  Sample B shows that for most applications, it is essential to avoid excessively rapid cooling of cuprous oxide crystals after growth or changes in the luminescence will occur.

Conduction carriers or other, nonthermalized states tend to be trapped at oxygen vacancies, while cold and diffusive excitons tend to excite copper vacancies.
The lifetime of the oxygen vacancy luminescence at low temperatures is no longer indicative of the exciton lifetime when copper vacancies are present because excitons are preferentially captured at copper vacancies.
When the time averaged luminescence of cuprous oxide shows primarily copper vacancy luminescence, for a brief period after pulsed excitation the luminescence is primarily from oxygen vacancies because they are pumped by conduction carriers.  Samples which show only oxygen vacancy luminescence are not necessarily n-type (oxygen deficient); instead, the process of copper vacancy excitation may be blocked by a high concentration oxygen vacancies.

Micro-optical density scale direct absorption by oxygen vacancies can be detected because the absorption is reduced when the oxygen vacancy is in an excited state.
There is a significant delay between bleach and luminescence in oxygen vacancies.
\section{Acknowledgements}
L.F. was supported by NSF IGERT DGE-0801685 and the Institute for Sustainability and Energy at Northwestern (ISEN).
        Crystal growth was supported by NSF DMR-1307698 and in part by Argonne National Laboratory under U.S. Department of Energy contract DE-AC02-06CH11357.
        K.C. was supported as part of the Center for Inverse Design, an Energy Frontier Research Center funded by the U.S. Department of Energy, Office of Science, Office of Basic Energy Sciences, under award number DE-AC36-08GO28308.
        Use of the Center for Nanoscale Materials was supported by the U. S. Department of Energy, Office of Science, Office of Basic Energy Sciences, under Contract No. DE-AC02-06CH11357.
        This work made use of the X-ray and OMM Facilities supported by the MRSEC program of the NSF DMR-1121262 at the MRC of Northwestern.
  \bibliographystyle{model1a-num-names} 
  \bibliography{vacancy}


\end{document}